\documentclass[preprint, review, 12pt]{elsarticle}

\usepackage{amssymb}
\usepackage{tasks}
\usepackage{subcaption}
\usepackage{graphicx}
\usepackage{xcolor}
\usepackage{multirow}
\usepackage{epstopdf}
\usepackage{url}
\usepackage[section]{placeins}
\usepackage[cmex10]{amsmath}
\usepackage[a4paper,left=3cm,right=3cm,top=3cm,bottom=3cm]{geometry}

\newlength\mylength
\begin{document}

\begin{frontmatter}



%\title{Calibrating Parameters of a Sub-transient Generator Model Using Dynamic Phasor Measurements}

\title{Estimation of the Plant Controller Communication Time-Delay Considering PMSG-Based Wind Turbines}


\author{Pablo Marchi$\dagger$, Pablo Gill Estevez$*$, Alejandro Otero$\dagger *$ and Cecilia Galarza$\dagger *$}

\address{\centering
$\dagger$ CSC-CONICET, Godoy Cruz 2390, Buenos Aires City, Argentina.\\
$*$ FIUBA, Paseo Col\'on 850, Buenos Aires City, Argentina.
}

\begin{abstract}
The communication control delay between the inverters and the power plant controller can be caused by several factors related to the communication link between them. Under undesirable conditions, high delay values can produce oscillations in the wind power plant that can affect the rest of the power system. In this work, we present a new robust methodology for wind turbines to estimate the value of the communication control delay using PMU data. Several scenarios are considered where external faults are simulated and the performance of the algorithm is evaluated based on dynamic state estimation of the mathematical model of the wind turbine. In this paper, we have shown that the characterization of the delay can be performed offering the transmission system operator an online tool to identify the most suited communication delay for the plant controller models used in dynamic studies.
\vspace{0.5cm}
\end{abstract}


\end{frontmatter}


\section{Introduction}

As the development and deployment of renewable energy sources increase, the relevance of wind turbine (WT) systems on power networks has become increasingly apparent \cite{Darwish2020}. On the other hand, the voltage stability problem of power systems with a high penetration of wind power gained more attention in the last decades \cite{jauch2006stability}. To maintain the point of common connection (PCC) voltage
stability of wind farms, a contribution of reactive power is necessary \cite{Liu2013}. Different mechanisms can be used to modify the reactive power of the wind power plant injected into the system. One of them, is to use a controller that modifies a reference value of the grid side converter using the voltage measured at the PCC \cite{Dai2022}. In this context, the robustness of the controller, against the communication delay between the control center and the turbines, is critical. The typical solution is to use a droop gain controller \cite{Asadollah2020}. However, this methodology presents some limitations when the communication delay is large. Besides, most wind power plants use centralized control and it is difficult to adjust when controller communication delays are present. In \cite{Cui2023}, a data-driven model estimation method for renewable energy sources (RES) is presented where random time delays are contemplated. However, this approach is focused on estimating the open-loop state matrix and the control input matrix of the system under normal operation of the wide-area control system.

Renewable plants are complex systems that require precise monitoring and control to optimize their performance. 
Dynamic state estimation (DSE) is a technique used to estimate the operating conditions and parameters of a wind turbine
in real-time based on available measurements. DSE plays a vital role for enhancing the performance, reliability, and maintenance of wind turbines. It enables operators to monitor the status of the turbine, detect anomalies, and take proactive measures to ensure safe and efficient operation \cite{HABIBI2019877}. By comparing the estimated states and parameters with their expected values or predefined thresholds, deviations, and faults should be detected. Advanced algorithms can analyze the discrepancies and provide insights into the type and severity of the detected faults. Besides, the estimated states and identified parameters can be utilized for turbine control and optimization strategies. By continuously updating the data obtained from different operating conditions of the turbines, control algorithms can make adjustments to maximize power generation, minimize loads, or improve system stability. A review of the published state estimation techniques for grid-connected wind turbine systems can be found in \cite{Mayilsamy2023}.

In \cite{Odgaard2013}, several types of fault are simulated to compare the performance of different strategies to solve the fault detection and isolation problem. Different fault classes were tested including false measurements from the sensors, errors in the actuators, and mechanical oscillations due to increased levels of drive train vibrations. Unfortunately, the authors did not perform mismatches between the simulations and the model used when communication delays are present. In \cite{SONG2017}, the authors proposed to use virtual measurements calculated from estimated states to optimize the baseline controller by enhancing the optimal tip speed ratio tracking and pitch adjustment. In this manner, the annual energy production can be approximately increased by $0.8\%$ compared to the baseline controller. In \cite{Afrasiabi2019} it has been shown that the ensemble Kalman filter (EnKF) is more accurate than the extended Kalman filter (EKF) and the unscented Kalman filter (UKF) to capture abrupt changes in states considering a full-detail model of a permanent magnetic synchronous generators based wind turbine (PMSG-WT). In \cite{Song2021}, a general dynamic state estimation framework for monitoring and controlling wind turbines is proposed showing that the DSE improves the accuracy of the state trajectory monitoring than the raw measurements. Here, the DSE framework can effectively monitor and control PMSG-WT even in the presence of noise and temporary natural or maliciously bad data. Unfortunately, this algorithm can not deal with unknown communication delays because this phenomenon remains longer than corrupted data over time.

\begin{figure}[]
\centering
\includegraphics[width = .8\columnwidth]{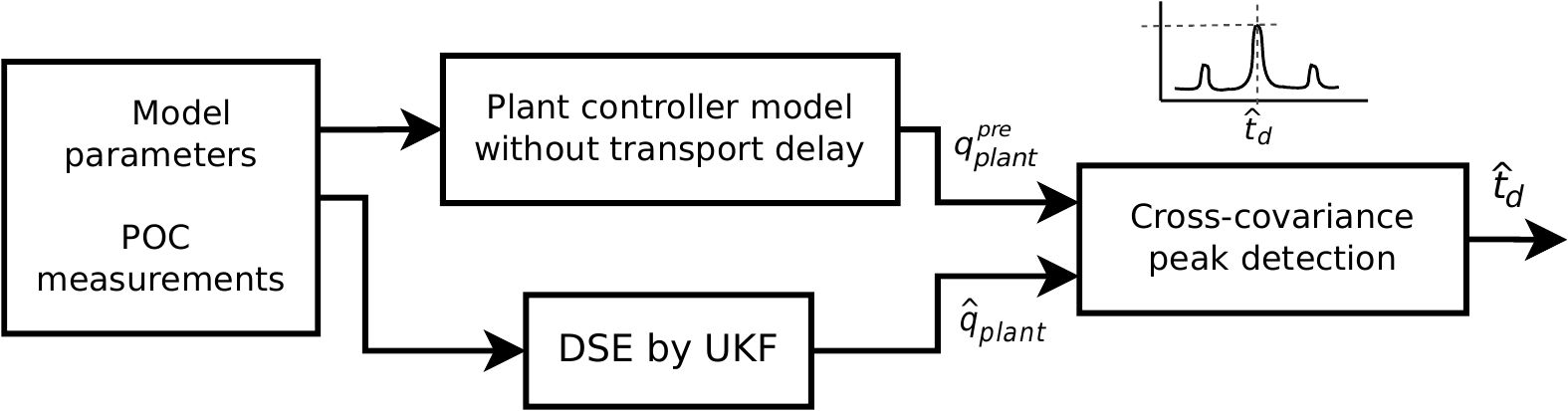}
\caption{Block diagram of the proposed methodology.}
\label{fig:proposed_methodology}
\end{figure}

In this paper, we present a new application of the DSE techniques to detect scenarios with delay in the communication link between the renewable energies and the control center. More specifically, we introduce an estimator that correctly estimates the internal states of a wind turbine even if oscillatory signals are present. Then, the estimated control signal is compared to the control system response without a transport delay. Finally, the delay time can be estimated using the cross-covariance value between these two signals. Fig. \ref{fig:proposed_methodology} shows a block diagram of our proposal. Our main contribution is to present a new methodology that is useful to adjust the model that transmission system operators use thereby obtaining realistic simulations in the dynamic system studies to be performed in the future. Besides, it also gives them the possibility to notify the wind farm owners if the delay exceeds the highest permitted value.

\section{Modeling Background} \label{sec:PMSG-WT model}

\begin{figure}[h]
\begin{centering}
\includegraphics[width=1\columnwidth]{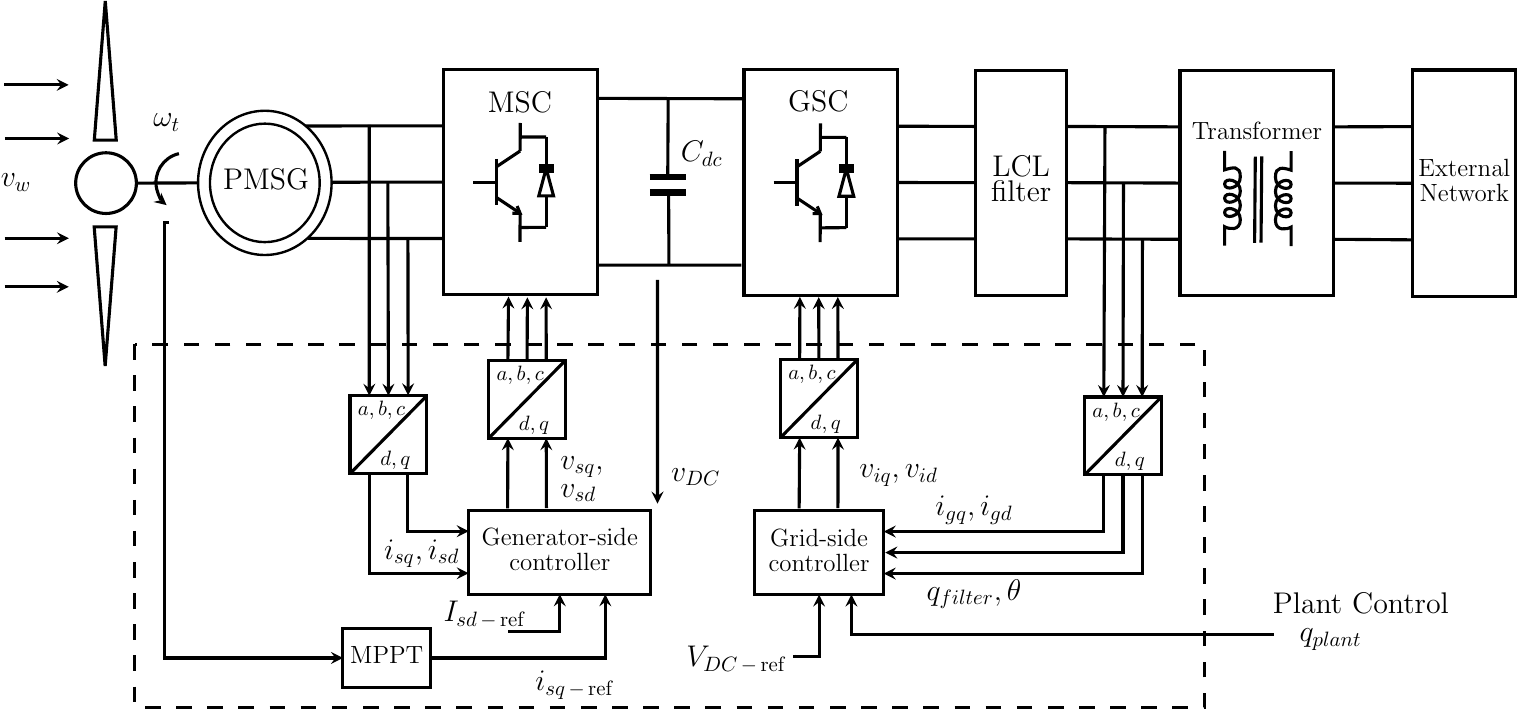}
\par\end{centering}
\caption{Schematic diagram of a PMSG-based WT.
.\label{fig:Plant-controller-breakdown-of}}
\end{figure}

%\subsection{PMSG-WT model} \label{sec:PMSG-WT model}
Our technique is applicable to several renewable energy plants. To make the development of our proposal clear, we have chosen the PMSG-WT as the renewable plant to be considered. PMSG-WT remain the preferred choice for applications demanding reliability \cite{Sunhua2022}. The main advantage of these turbines is their high efficiency. PMSGs have higher efficiency compared to traditional induction generators. The absence of rotor copper losses and the high
magnetic field strength of the permanent magnets contribute to improved power conversion efficiency \cite{Yang2012}. A block diagram of the different parts and controllers of PMSG-WTs
can be observed in Fig. \ref{fig:Plant-controller-breakdown-of}. These WTs employ a permanent magnet synchronous generator. The permanent magnets generate a magnetic field. As the rotor turns, it induces a changing magnetic field in the stator coils, creating an electric current. Besides, the generator is decoupled from the rest of the power grid using a machine-side converter (MSC) and a grid-side converter (GSC). For this reason, these WTs are usually called full converter WTs. In addition, an LCL filter is used to achieve high levels of harmonic attenuation. 

\subsection{Aerodynamic and turbine model} 
The most common assumption used for aerodynamic modeling is to consider a linear relationship between the turbine power and the wind power $p_{turb} = c_p \, p_w$, where $c_p$ is the performance coefficient representing the portion of wind power that is extracted by the turbine. Then, $p_w$ can be computed as the rate of change of the kinetic energy of a mass of air moving at a speed $v_w$. So, replacing $p_w$ by its respective value the following equation is obtained \cite{Wu2011}: 
\begin{equation} 
p_{turb} = 0.5 \, \rho \, \pi \, B_l^2 \, c_p(\lambda, \beta) \, v^3_w 
\end{equation}

\begin{figure}[]
\begin{centering}
\includegraphics[width=.95\columnwidth]{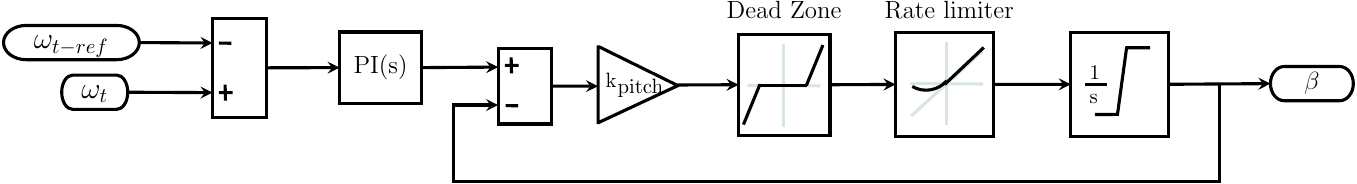}
\par\end{centering}
\caption{Block diagram of the pitch controller.\label{fig:pitch-controller}}
\end{figure}

Here, $B_l$ is the length of the blade \footnote{In this article, constants and time-varying signals are defined using uppercase and lowercase letters respectively.}, and $\rho$ is the air density. Generally, for power system studies $c_p$ is approximated and computed as a function of the tip-speed ratio  $\lambda = \omega_t \,B_l / v_w$ and the pitch angle $\beta$ , where $\omega_t$ is the turbine angular velocity. Besides, the pitch angle $\beta$ is changing considering the pitch controller in Fig. \ref{fig:pitch-controller}. On the other hand, the turbine generator mechanical system is represented using a single mass-spring-damper model as follows:
 \begin{equation} 
\frac{d\, \omega_t}{dt} = \frac{1}{2H} \left( t_t - t_g \right),
\label{eq:turbine_model} 
\end{equation}
where $t_t$ is the turbine torque, which is computed as \mbox{$t_t = p_{turb}/, \omega_t$}, $H$ is the inertia constant and $t_g$ is the generator electromagnetic torque. 

\subsection{PMSG model} 

Using a standard dq-coordinate system, the electrical equations for a synchronous generator can be written as \cite{kundur1994power}: 
\begin{subequations}
\begin{align}
    \frac{d\, i_{sq}}{dt} &= \frac{\omega_{elB}}{L_q} \left(-R_a\,i_{sq} - v_{sq} - L_d\, i_{sd}\, \omega_t +  \Phi_{pm}\, \omega_t \right) \\
    \frac{d\, i_{sd}}{dt} &= \frac{\omega_{elB}}{L_d} \left(-R_a\,i_{sd} - v_{sd} + L_q\, i_{sq}\, \omega_t \right)
\end{align}
\end{subequations}

where $v_{s}$, $i_{s}$, $\Phi_{pm}$, $\omega_{elB}$, $L_d/L_q$, $R_a$ represent the stator voltage, the stator current, the permanent magnet flux, stator inductance in $d/q$ axis and stator resistance respectively.

\subsection{MSC controller} 
The MSC in PMSG uses a decoupling control strategy where the d-axis regulates reactive power and the q-axis regulates electrical torque in order to ensure the maximum power operation point. To illustrate our analysis, we have chosen the model used in \cite{kunjumuhammed2019}:
\begin{subequations}
\begin{align}
    \frac{d\, v_{q\,MSC}}{dt} &= K_i{}_{\,MSC-IL1} \, \left( \omega_t^2 \, \frac{K_{opt}}{\Phi_{pm}} - i_{sq}  \right) \\
    \frac{d\, v_{d\,MSC}}{dt} &= K_i{}_{\,MSC-IL2} \,\, \left( I_{sd - ref} - i_{sd}  \right) \label{eq:sec_MSC}
\end{align}
\end{subequations}

where $K_i{}_{\, \left\lbrace \cdot \right\rbrace}$ are the constants for the integrator controller, $I_{sd - ref}$ is the reference value used to adjust the direct current of the generator, and 
$K_{opt}$ is obtained from the maximum power operating point.

\subsection{Back-to-back capacitor and GSC controller}

The back-to-back capacitor is the core of the DC link which is a section of the 
electrical system that smoothens and stabilizes the electrical output from the 
generator. It is modeled as a simple capacitor as:
\begin{subequations}
\begin{align}
\frac{d\,v_{DC}}{dt} &= \frac{1}{C_{dc}\,v_{DC}} \left( p_{pmsg} - p_{gsc} \right) 
\end{align}
\end{subequations}

where $v_{DC}$ is the capacitor voltage, $C_{dc}$ is the DC-link capacitance, $p_{pmsg}$ the generator  active power output, and $p_{gsc}$ the GSC output active power. 

Similarly to the MSC, the GSC uses a decoupled independent control for the d-q axis. The developed generator model assumes that the q-axis is aligned with the slack bus voltage defined in the power flow. Therefore, these voltages and currents must be rotated to the axis aligned to the WT terminal bus voltage. So, a phase-locked loop (PLL) is used to find the required angle of rotation. However, in this paper, the PLL is not included, and non-modeling errors in the terminal voltage angle $\theta$ are contemplated, as in \cite{Anagnostou2019}. The equations of the GSC model are defined below:
\begin{subequations}
\begin{align}
\frac{d\, i_{q\,GSC}}{dt} & = K_i{}_{\,GSC-OL1} \left(V_{DC-ref}-v_{DC} \right) \\
\frac{d\, v_{q\,GSC}}{dt} & = K_i{}_{\,GSC-IL1} \, \left[ i_{q\,GSC} + K_p{}_{\,GSC-OL1} \right. \nonumber \\
 & \qquad \qquad \qquad \left. \left(V_{DC-ref}-v_{DC} \right) - i'_{gq} \right] \\
\frac{d\, i_{d\,GSC}}{dt} & = K_i{}_{\,GSC-OL2} \left(q_{plant}-q_{filter} \right) \\
\frac{d\, v_{d\,GSC}}{dt} & = K_i{}_{\,GSC-IL2} \, \left[ i_{d\,GSC} + K_p{}_{\,GSC-OL2} \right. \nonumber \\
 & \qquad \qquad \qquad \left. \left( q_{plant}-q_{filter} \right) - i'_{gd} \right]
\label{eq:GSC_controller}
\end{align}
\end{subequations}

where $K_p{}_{\, \left\lbrace \cdot \right\rbrace}$ are the constants for the proportional controllers, $V_{DC-ref}$ is the reference voltage used for the DC-link, $i_{q\,GSC}/
i_{d\,GSC}$ are the state variables related to the DC-link capacitor voltage PI 
control, $q_{plant}$ is the signal used by the plant controller to modify the 
reactive power of the WT usually defined by the wind farm operator, $q_{filter}$ is the reactive output (it can be computed as $q_{filter} = v_{gd} \, i_{gq} - v_{gq} \, i_{gd}$), and $i'_{gq}/
i'_{gd}$ are the terminal currents in $d/q$ axis \footnote{The notation $'$ 
indicates that the variable is computed considering the terminal voltage rotating 
reference frame $\theta$. For example, $i_{gq}+\,\mathrm{i}\,i_{gd} = \left( i'_{gq}+\,\mathrm{i}\,i'_{gd} 
\right) e^{\mathrm{i} \theta}$, where $\mathrm{i}$ is the imaginay unit.}  

\subsection{LCL filter} 

The purpose of adding an additional filter is to remove switching harmonics. The 
complete set of equations of this stage are \cite{kunjumuhammed2019}: 
\begin{subequations}
\begin{align}
\frac{d\, v_{cq}}{dt} &= \frac{\omega_{elB}}{C_f} \left[ i_{iq} - i_{gq} +  C_f\, v_{cd}   \right] \label{eq:LCL_filter_first} \\
\frac{d\, v_{cd}}{dt} &= \frac{\omega_{elB}}{C_f} \left[ i_{id} - i_{gd} -  C_f\, v_{cq}   \right] \\
\frac{d\, i_{iq}}{dt} &= \frac{\omega_{elB}}{L_i} \left[ v_{iq} - v_{cq} - (R_i + R_c)\, i_{iq} +  L_i\, i_{id} + R_c\, i_{gq} \right] \label{eq:LCL_filter_2} \\
\frac{d\, i_{id}}{dt} &= \frac{\omega_{elB}}{L_i} \left[ v_{id} - v_{cd} - (R_i + R_c)\, i_{id} -  L_i\, i_{iq} + R_c\, i_{gd} \right]   \label{eq:LCL_filter_3}\\
\frac{d\, i_{gq}}{dt} &= \frac{\omega_{elB}}{L_g} \left[ v_{cq} - v_{gq} - (R_g + R_c)\, i_{gq} +  L_g\, i_{gd} + R_c\, i_{iq} \right] \\
\frac{d\, i_{gd}}{dt} &= \frac{\omega_{elB}}{L_g} \left[ v_{cd} - v_{gd} - (R_g + R_c)\, i_{gd} -  L_g\, i_{gq} + R_c\, i_{id} \right] 
\label{eq:LCL_filter_end}
\end{align}
\end{subequations}
Here, $R{}_{\, \left\lbrace \cdot \right\rbrace}$, $L{}_{\, \left\lbrace \cdot \right\rbrace}$ and $C{}_{\, \left\lbrace \cdot \right\rbrace}$ represent the filter resistances, inductances, and 
capacitances respectively, $v_{cq}/v_{cd}/v_{iq}/v_{id}$ $i_{iq}/i_{id}$ are 
filter voltages and currents in the $d/q$ axis, and $\omega_{elB}$ is the angular 
velocity base.
\subsection{Power plant controller}
\begin{figure}[]
\begin{centering}
\includegraphics[width=1\columnwidth]{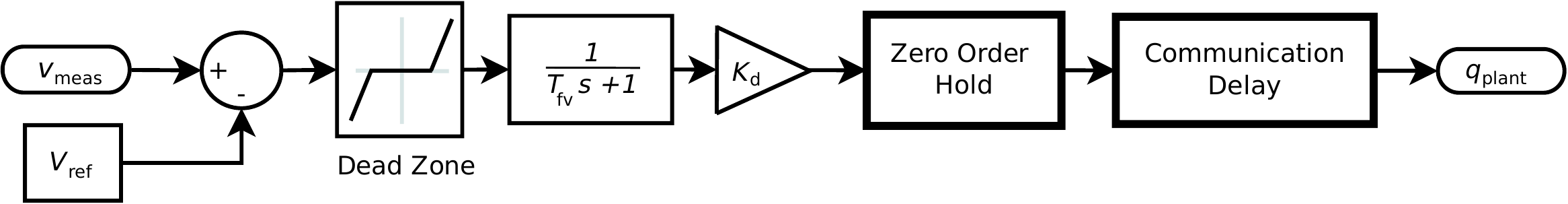}
\par\end{centering}
\caption{Block diagram of the plant controller. \label{fig:plant-controller}}
\end{figure}

As it was mentioned in the introduction, the reactive controller in wind plants plays a vital role in ensuring voltage stability, managing reactive power, complying with grid codes, and responding dynamically to grid conditions. In this paper, we considered a PI controller as it is shown in Fig. \ref{fig:plant-controller}. The measured terminal absolute value of the voltage at PCC $v_{meas}$ is compared to a parametric reference $V_{ref}$ to define the reactive power $q_{plant}$ that will be used by the GSC of the turbine generators. A simplified version of the model used is represented by \eqref{eq:plant controller}, where $dz$ represents the dead zone function. However, as in \cite{Fan2023}, we also considered the discretization errors introduced in the signal, modeled as a zero-order hold, and the communication delay.
\begin{equation}
\frac{d \,q_{plant}}{dt} = \frac{1}{T_{fv}} \, \left[ \, dz(v_{meas} - V_{ref}) \, K_d - q_{plant} \right]
\label{eq:plant controller}
\end{equation}

\vspace{0.5cm}

\section{Proposed Methodology} \label{sec:Method}

If the WT variables used in equations \eqref{eq:turbine_model}-
\eqref{eq:plant controller} are grouped into a state vector, the following general form a 
discrete-time state-space model for nonlinear systems can be used:
\begin{equation}
\begin{aligned}
\mathbf{x}_{k} &= f_s(\mathbf{x}_{k-1},\mathbf{u}_{k-1}) + \mathbf{w}_{k} \label{eq:nonlinearsystem1}\\
\mathbf{z}_{k} &= h_s(\mathbf{x}_{k},\mathbf{u}_{k})+\mathbf{v}_{k}, 
\end{aligned}
\end{equation}

where $k$ is the discrete time, $f_s$ is the state transition function that models the system dynamics, 
$h_s$ is the measurement function which relates the measurements with the state 
vector, $\mathbf{x}\in\mathbb{R}^{n \times 1}, \mathbf{z}\in\mathbb{R}^{m \times 
1}, \mathbf{u}\in\mathbb{R}^{l \times 
1}$ are the state/measurement/input vectors respectively, and 
$\mathbf{w}\in\mathbb{R}^{n \times 1}$ and $\mathbf{v}\in\mathbb{R}^{m \times 1}$ 
are noise vectors introduced to contemplate modeling errors.

%All the notation used for this section can be found in Appendix \ref{sec:nomenclature}.

To estimate the state vector of a WT system, several Bayesian filters can be 
implemented. Among them, the EKF, UKF, and EnKF are the most common options 
\cite{Mayilsamy2023}. The decision of which algorithm to be used depends on the 
problem to be solved. Indeed, there is a trade-off between the computational 
burden and the accuracy of the algorithm in the presence of non-linearities. 
Generally, EKF performs well when the system model presents low linearization errors. For this algorithm, the jacobians of the $f_s$ and $h_s$ functions must be 
computed but the computational cost is relatively low. If the jacobians are  
laborious to obtain and the functions are highly non-linear, the UKF and the EnKF 
are more suitable alternatives.

In this paper, a reduction of the model presented in Section \ref{sec:PMSG-WT model} is performed to decrease the complexity and to ensure the observability of the system. Accordingly, and as it was considered in \cite{Anagnostou2019}, the state variables related to the LCL filter will be considered as algebraic variables, defined as $\mathbf{av}=\left[ v_{cq},\,v_{cd},\,i_{iq},\,i_{id}, \,i_{gq},\,i_{gd} \right]^T$, because their dynamics are characterized by very low time constants compared
to the simulation time step. For example, considering the values used in \cite{kunjumuhammed2019}, $L_g/\omega_{elB} = 8.75 \, \mu$s, which is the time constant of $i_{gq}$ and $i_{gd}$, is too low compared with the reporting rates of the PMUs (measurements in ms). On the other hand, the DC voltage of the link between the MSC and the GSC is considered as a constant, i.e., the variable $v_{DC}$ is equal to $V_{DC-ref}$. With these considerations, the equations related to the aerodynamic, turbine, PMSG, MSC, and back-to-back capacitor can be omitted. Besides, with this approximation, we guarantee system observability using the measurements at PCC uniquely. This fact is a significative difference to the rest of the papers included in the literature, for example, as in \cite{Song2021}.

So, the equations used to compute the algebraic variables are obtained by considering that the derivative of each variable is equal to zero.
\begin{equation}
g_s(\mathbf{x},\mathbf{u},\mathbf{av})=\mathbf{0},
\label{ecu:algebraic_state_variables}
\end{equation}

where $g_s$ is a set of implicit functions defined by equations \eqref{eq:LCL_filter_first} - \eqref{eq:LCL_filter_end}. In this context, the system model described by \eqref{eq:nonlinearsystem1} is used by the Bayesian filter. As a consequence of the approximations explained before, the state vector $\mathbf{x}\,\in\,\mathbb{R}^{^{6x1}}$ is defined using the internal states of the GSC and the reactive power controller signal as:
\begin{equation}
\mathbf{x}=\left[i_{q\,GSC},\,v_{q\,GSC},\,i_{d\,GSC},\,v_{d\,GSC}, \, q_{plant},\, \sigma_{i_{d\,GSC}} \right]^T,
\end{equation}
This definition focuses on the reactive control signal and the variables that affect the reactive output of the WT because the real-time estimation of the plant controller signal is necessary to reach the main objective of this paper. Here, the $\sigma_{i_{d\,GSC}}$ is the bias of the $i_{d\,GSC}$ variable and it was added for practical purposes, because we have noticed that the state $i_{d\,GSC}$ had a bias that must be corrected.

Then, 
$\mathbf{u}$ and $\mathbf{z}$ are 
defined using voltage and current phasors which 
are acquired through intelligent electronic 
devices, such as PMUs, connected at PCC. The measurement vector is defined as the $d/q$
terminal voltages as :
\begin{equation}
\mathbf{z}=\left[v_{gq},\,v_{gd} \right]^T
\end{equation}

Besides, the input vector ($\mathbf{u}$) includes the terminal voltage angle, voltages and currents. It can be defined as:
\begin{equation}
\mathbf{u}=\left[\theta,\,v_{gq},\,v_{gd},\,i_{gq},\,i_{gd} \right]^T
\end{equation}

%the measurement function depends on state and algebraic variables $h_s(\mathbf{x},\mathbf{av})$ 
The state transition function $f_s(\mathbf{x},\mathbf{u})\,\,\in\,\left(\mathbb{R}^{^{6x1}},\mathbb{R}^{^{5x1}}\right)\rightarrow\mathbb{R}^{^{6x1}}$ is defined in \eqref{eq:transition_function}, where $\Delta t$ is the step time. 

\begin{equation}
f_s(\mathbf{x},\mathbf{u})= \mathbf{x}\, +\, \left[\begin{array}{c}
0\,\\
K_{i_{GSC\,IL1}}\left(i_{q\,GSC}\,-\,i'_{gq}\right)\,\\
K_{i_{GSC\,OL2}}\left(q_{plant}-q_{filter}\right) - \sigma_{i_{d\,GSC}}\,\\
K_{i_{GSC\,IL2}}\left[i_{d\,GSC}+K_{p_{GSC\,OL2}}\left(q_{plant}-q_{filter}\right)-i'_{gd}\right]\\
\frac{1}{T_{fv}} \, \left[ dz(V_{term} - V_{term_{ref}}) \, K_d - q_{plant} \right] \\
0
\end{array}\right] \quad \Delta t,
\label{eq:transition_function}
\end{equation}

The role of the measurement function is to define a relationship between the measurements and the state variables to properly implement the correction stage of the algorithm used. Thus, it is of vital importance to define $h_s(\mathbf{x},\mathbf{u})$ using the algebraic variables and to avoid using the input values uniquely. It is worth mentioning that the algebraic variables, input vector, and state vector are correlated by \eqref{ecu:algebraic_state_variables}. As the algebraic variables depend on the state variables, when a prediction of the state is made it affects the evolution of the algebraic variables as well. 
So, the complete formulation of $h_s(\mathbf{x}_{k},\mathbf{u}_{k})$ requires the computation of $\mathbf{av}$. It is too extensive to be included in the manuscript but we will mention the steps to calculate the measurement vector.

First, the variables $i'_{gq}$, $i'_{gd}$ and $q_{filter}$ are computed using $\mathbf{u}$ and non-linear functions. Then, the variables $v_{iq}$ and $v_{id}$, present in \eqref{eq:LCL_filter_2} and \eqref{eq:LCL_filter_3}, are computed using:

\begin{subequations}
\begin{align}
v'_{iq} &= v_{q\,GSC} + K_p{}_{\,GSC-IL1} \left( i_{q\,GSC} - i'_{gq} \right) \\
v'_{id} &= v_{d\,GSC} + K_p{}_{\,GSC-IL2} \left[ i_{d\,GSC} + K_p{}_{\,GSC-OL2} \right. \nonumber \\
 & \qquad \qquad \qquad \left.\left( q_{plant}-q_{filter} \right) - i'_{gd} \right] 
\end{align}
\end{subequations}

and a new reference frame transformation. Finally, to obtain $h_s(\mathbf{x},\mathbf{u})$,  a linear subset of six equations with six algebraic variables \mbox{$\mathbf{av}^* = \left[ v_{cq},\,v_{cd},\,i_{iq},\,i_{id}, \,v_{gq},\,v_{gd}\right]$} is solved, using equations \eqref{eq:LCL_filter_first}-\eqref{eq:LCL_filter_end}. 

% As the oscillatory scenarios can be caused by perturbations in the reference value of the converters \cite{Fan2023}.

Given the need to deal with the non-linear functions $f_s$ and $h_s$ and after performing a reduction of TG model, we propose to use the sigma-point-based Kalman filtering (UKF) \cite{Wan2000}. It was originally used for synchronous generators and adapted to WT models later. The UKF linearizes the system dynamics, considering Gaussian 
assumptions, using a 
deterministic set of points called sigma points. Through these sigma points, the mean and the covariance of the 
predicted state vector can be obtained. The sigma points in the UKF are computed using the following equation: 
\begin{align}
\hat{\mathbf{x}}_{k-1}^{j} & =\hat{\mathbf{x}}_{k-1}+\left[\sqrt{\left(n+\lambda\right)P_{k-1}}\right]_{j}\, \qquad \hat{\mathbf{x}}_{k-1}^{0} =\hat{\mathbf{x}}_{k-1},\\
\hat{\mathbf{x}}_{k-1}^{j+n} & =\hat{\mathbf{x}}_{k-1}\,-\left[\sqrt{\left(n+\lambda\right)P_{k-1}}\right]_{j}\,,\quad j=1,\cdots,n,\nonumber
\label{eq:UKF1}
\end{align} 

where \mbox{$\lambda =\gamma^2(n+\kappa)-n$}, $\gamma,\, \kappa$ are positive scalars , $\left[ \, \cdot \right]_j$ is an operator that takes the $j$-th row of the matrix. Here the number of sigma points is restricted by the number of state variables $n$. Then the estimation process is obtained after the prediction and update steps are performed consecutively as follows:\footnote{Here, $W_m$ and  $W_c$ are the weights of the mean and covariance respectively \cite{Wan2000}.}:

\subsubsection*{State prediction:}
\begin{equation}
\begin{aligned}
\mathcal{X}^{(l)}_k &= f_s(\hat{\mathbf{x}}_{k-1}^{l}, \mathbf{u}_{k-1}), \quad \hat{\mathbf{x}}_{k} = \sum_{l=0}^{2n} W_m^{l}\, \mathcal{X}^{(l)}_k \\
P_k &=  \sum_{l=0}^{2n} W_c^{l} \, \left( \mathcal{X}^{(l)}_k -  \hat{\mathbf{x}}_{k} \right) \left( \mathcal{X}^{(l)}_k -  \hat{\mathbf{x}}_{k} \right)^T + Q_k
\end{aligned} 
\end{equation}

\subsubsection*{Measurement update of the state:}
\begin{equation}
\begin{aligned}
\mathcal{Z}^{(l)}_k &= h_s(\hat{\mathbf{x}}_{k-1}^{l}, \mathbf{u}_{k}), \quad \hat{\mathbf{z}}_{k} = \sum_{l=0}^{2n} W_m^{l}\, \mathcal{Z}^{(l)}_k \\
P_{z\,k} &= \sum_{l=0}^{2n} W_c^{l} \, \left( \mathcal{Z}^{(l)}_k -  \hat{\mathbf{z}}_{k} \right) \left( \mathcal{Z}^{(l)}_k -  \hat{\mathbf{z}}_{k} \right)^T + R_k \\
P_{xz\,k} &= \sum_{l=0}^{2n} W_c^{l} \, \left( \mathcal{X}^{(l)}_k -  \hat{\mathbf{x}}_{k} \right) \left( \mathcal{Z}^{(l)}_k -  \hat{\mathbf{z}}_{k} \right)^T
\end{aligned}
\end{equation}
\begin{equation}
\begin{aligned} 
\hat{\mathbf{x}}_{k} &= \hat{\mathbf{x}}_{k} + K_k (\mathbf{z}_{k} - \hat{\mathbf{z}}_{k}), \quad P_k = P_{k} - K_k P_{z\,k} K_k^T\\
K_k &= P_{xz\,k}\,P_{z\,k}^{-1}
\end{aligned}
\vspace*{0,2cm}
\end{equation}

After performing the DSE, and obtaining an estimated state vector $\hat{\mathbf{x}}_{k}$,  an estimated $\hat{q}_{plant}$ can be compared with a prediction $q^{pre}_{plant}$ using only the model defined by \eqref{eq:plant controller}. As this model does not include the transport delay of the $q_{plant}$ signal, $q^{pre}_{plant}$ can differ from the estimated $\hat{q}_{plant}$. Another difference between the signals is that $\hat{q}_{plant}$ is obtained from the UKF, so it has been filtered and contains much less high-frequency components than $q^{pre}_{plant}$. However, as the main difference between the signals is the transport delay, the cross-covariance function \cite{Kun2018} can be used to determine the value of the delay between these two signals. The location of the maximum of the cross-covariance function indicates the most probable time shift. Thus, the delay is computed through a peak detection algorithm.

\section{Numerical Results}
\begin{figure}[]
\centering

\includegraphics[width = .9\columnwidth]{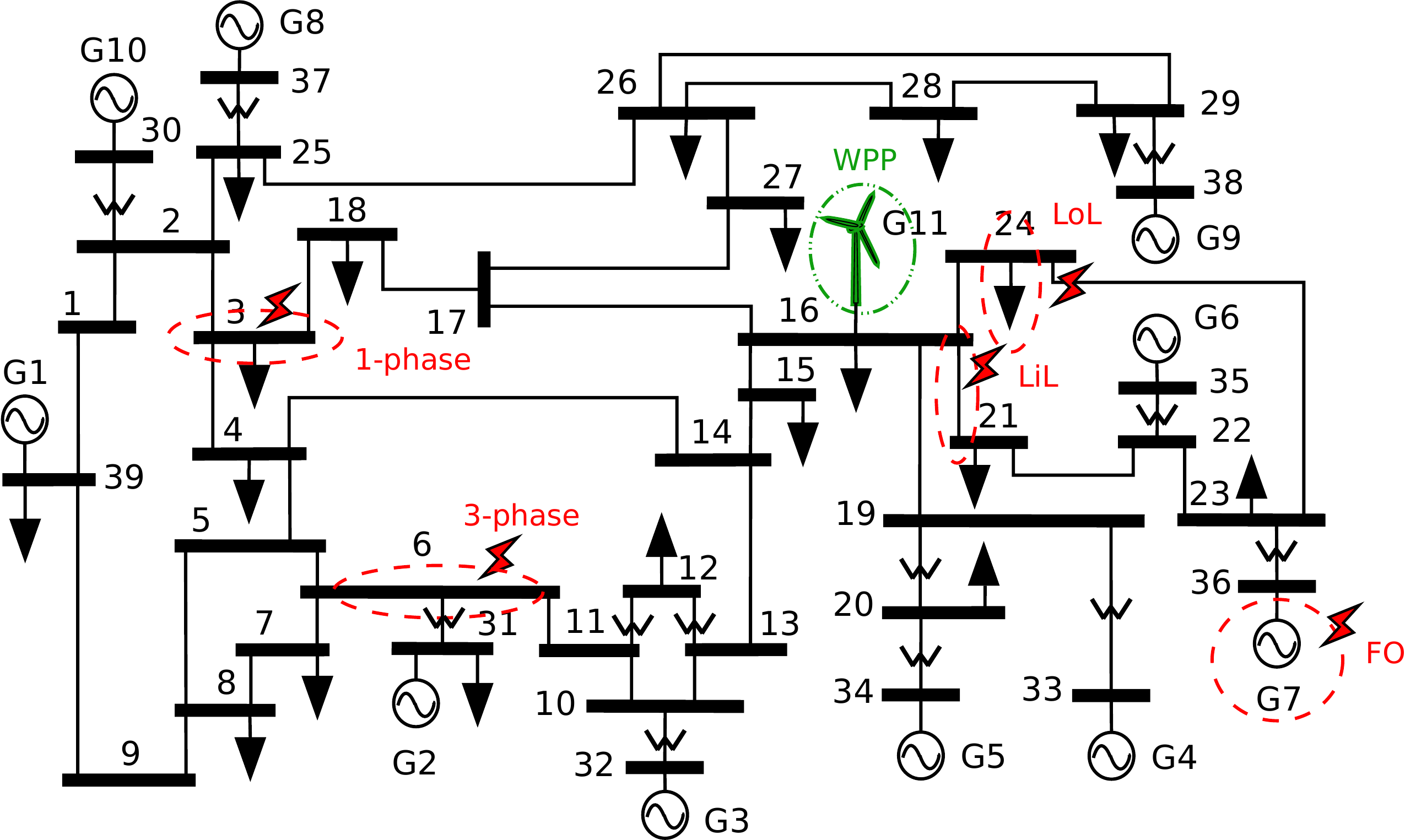}
\caption{IEEE 39 bus system.}
\label{fig:39_bus}
\end{figure}

In this section, all the simulations were performed using MATLAB \& Simulink software and considering the IEEE 39-bus system presented in \mbox{Fig 
\ref{fig:39_bus}}. The IEEE 39-bus standard system is a power network in the New England area of the United States. The system consists of 10 generators, 39 busbars, and 12 transformers. The models used for the generators, transformers, lines, and loads, and the complete set of parameters can be found in \cite{Patrice2016}. The simulated system can be used for traditional transient stability analysis, excitation, speed controller design, system frequency modulation, and other characteristic analysis. 

To make the analysis more attractive, instead of simulating a unique WT, an equivalent of a wind power plant (WPP) is installed in bus number 16 as G11. We use the “1+1” model that takes the equivalent
expression of the wind farm as one wind turbine and one
generator \cite{ZOU2014956}. The model used for the power plant equivalent is the same as the model described in section \ref{sec:PMSG-WT model}, except for the base power, which is computed as the sum of the power of each turbine, and the equivalent transformer and line at PCC \cite{ZOU2014956}. Besides, the wind speed is the average of the wind speeds for each generator. In this paper, the model represents a wind farm of 29 turbines with a nominal power of $P_{turb} = 2$ MW each. The block diagram and the parameters of the model used can be found in \cite{kunjumuhammed2019}, including the control loop for the pitch angle presented in  \cite{kunjumuhammed2019} that it was not described in section \ref{sec:PMSG-WT model}. In the following subsections, several case scenarios will be described where different perturbations to the system are applied singly. In this context, and in order to evaluate the behavior of the methodology, some external faults have been simulated. It is important to mention that these external faults generate a significant variation of the $q_{plant}$ variable allowing the proper functioning of our proposal. Nevertheless, when the system is operating at a stationary point the variations of the $q_{plant}$ are unnoticeable, and the cross-covariance can detect incorrect delay times due to the similarity of the signal in different instants of time.

\subsection{3-phase to ground fault (3-P)}
This first scenario (case 3-P) simulates a three-phase to-ground fault in \mbox{bus 6}. This case is considered because it is a fault relatively far from the point of connection of the WPP (bus 16). The fault is simulated at time $t_{fault}  = 5$s and remains active for $0.1$s, after that it is cleared. 

%\section{Discussion}
%In Fig \ref{fig:CaseA_residuals}, the residuals are shown. As it can be observed, the mean of the residuals is close to zero which suggests that the filter is correctly estimating the state variables.

\subsection{1-phase to ground fault (1-P)}
In this case (case 1-P) a single-phase to-ground fault \mbox{in bus 3} is simulated. As in the previous case, the fault is simulated at time $t_{fault}  = 5$s and remains active for $0.1$s, after that it is cleared. This is one of the most common faults that can occur in a power system and it was located at a bus closer to the WPP but still at a considerable distance.

\subsection{Forced oscillation in G7 (FO)}
The third scenario (case FO) simulates a forced oscillation in G7 that is propagated by the rest of the network. In this case, an oscillatory component is added to the mechanical power signal of G7 with an oscillation frequency $f_{osc}=0.1$Hz and an amplitude $a_{osc} = 0.16$p.u. This scenario is included to contemplate cases where the signals at the point of connection are constantly changing. In this case the $q_{plant}$ signal oscillates according to the variations of the changes in the terminal voltage.

\subsection{Load loss (LoL)}
In this case, the load loss located at bus 24 is simulated. This scenario (LoL), represents a possible blackout generated by a fault in the respective bus. This event affects the rest of the system and generates a big change in the reactive power delivered by the WPP because of the proximity between buses.

\subsection{Line loss (LiL)}
The following case is the loss of the line between the bus 16 and 21. It is refered as line loss (LL) and it an event that produces a slight perturbation of the terminal voltage, making that the $q_{plant}$ variable changes from a satationary point to another following an oscillatory behaviour.

\vspace{0.5cm}
In all cases, the final time of the simulation is set to 60 s and the reporting rate of the PMU to $60$ fps ($\Delta t = 1/60 \approx  0.0167 \, \mathrm{s}$), so it results in $N_{samp} = 3600$, where $N_{samp}$ is the number of samples used in the simulation. As different levels of noise can be present in the measurements, and in the input variables of the algorithm, the methodology is evaluated considering three values of the total vector error (TVE). Here, an additive white Gaussian noise is used producing a $TVE= \left\lbrace 3, 1, 0.1 \right\rbrace$ \%. Notice that the first two values that were chosen are consistent with the worst scenarios of the synchrophasor standard \cite{IECStandard2018} for dynamic and stationary conditions respectively. Besides, the time delays to be analyzed are $t_d = \left\lbrace 50, 100, 500 \right\rbrace$ ms because as the nature of the delay is a systematic error we do not consider that it should be treated as a random variable. The first two values correspond to real practical cases and the last one is related to cases of abnormal functioning.

\begin{figure}
     \centering
     \begin{subfigure}[b]{1\textwidth}
         \centering
				 \caption{Estimated $q_{plant}$ signals in per unit.}
         \includegraphics[width=1\columnwidth]{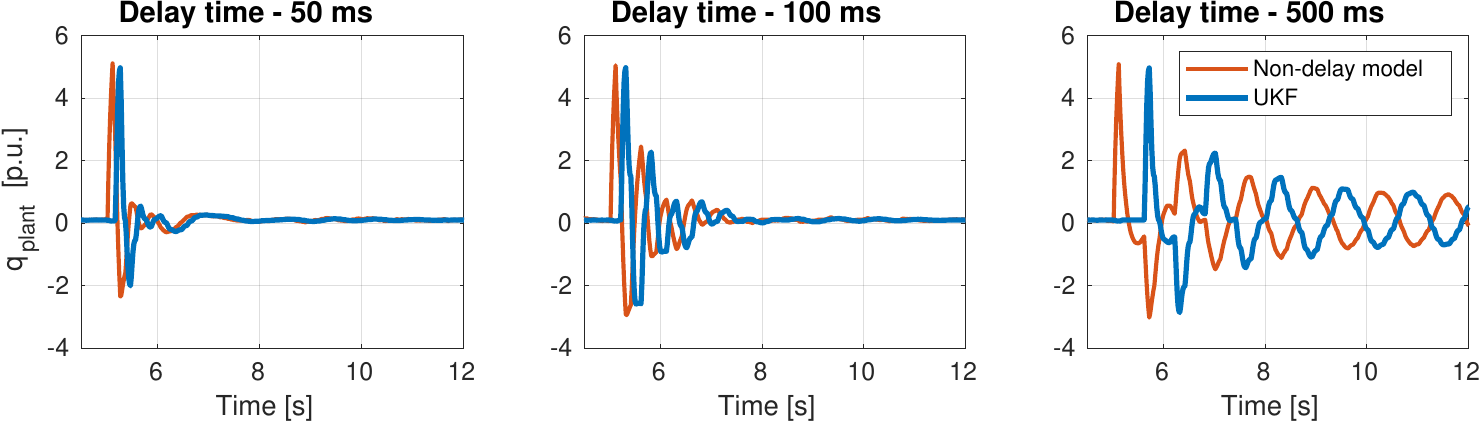}
         \label{fig:signals_delay_comparison}
     \end{subfigure}
     \begin{subfigure}[b]{1\textwidth}
         \centering
         \caption{Cross-covariance between the $\hat{q}_{plant}$ signals.}
 				 \includegraphics[width=1\columnwidth]{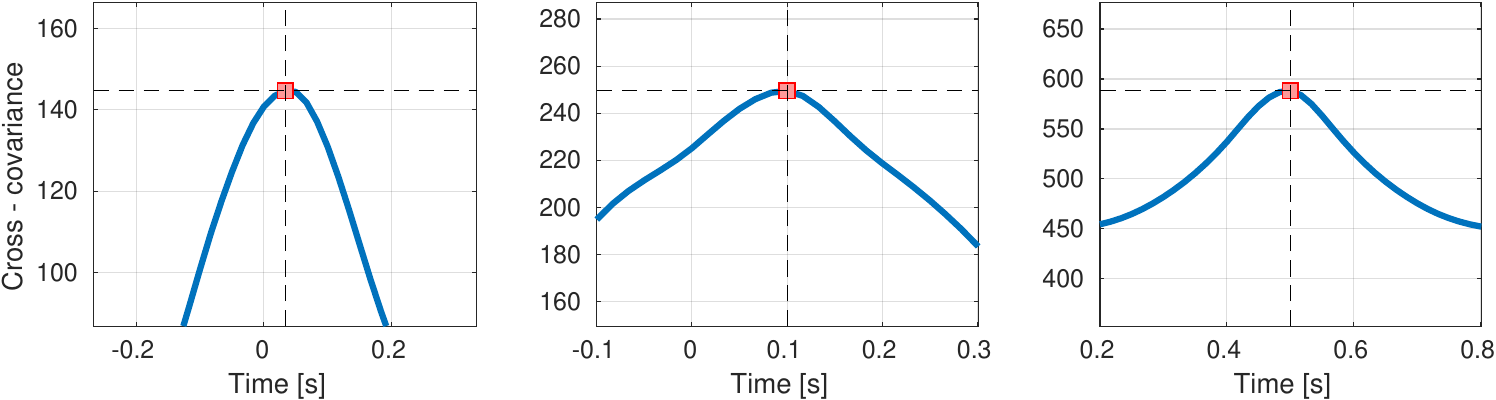}	 
				 \label{fig:crosscov_delay_comparison}
     \end{subfigure}
     \caption{Cross-covariance results in case 3-P and $TVE=1$ \%.}
		 \label{fig:delay_comparison}
\end{figure}

In Fig. \ref{fig:signals_delay_comparison}, a comparison between $\hat{q}_{plant}$, obtained by the UKF, and $q_{plant}^{pre}$, obtained using \eqref{eq:plant controller}, is presented. As it can be clearly observed, the delay produced by the communication link is included in the $\hat{q}_{plant}$ signal. The main reason for this result is that the UKF obtains $\hat{q}_{plant}$ that reduces the mean squared error between the estimated state vector and the true one. So, the measurements are used as feedback to correct the value of the states instead of using an open loop as it is done in \eqref{eq:plant controller}. Indeed, the $q_{plant}^{pre}$ is a noisy and a time-shifted version of $q_{plant}$, where the impact of the noise is higher, because there is not a filter stage involved for its computation. Using $\hat{q}_{plant}$ and $q_{plant}^{pre}$ we compute the cross-covariance function. Fig. \ref{fig:crosscov_delay_comparison} shows the different cross-covariance functions. It is clear that the maximum value of this function is located at different values of time according to the delay communication time simulated. The degree of accuracy is remarkable in these cases.

\begin{figure}
     \centering
     \begin{subfigure}[b]{1\textwidth}
         \centering
         \caption{$TVE=0.1 \%$.}
         \includegraphics[width=.9\columnwidth]{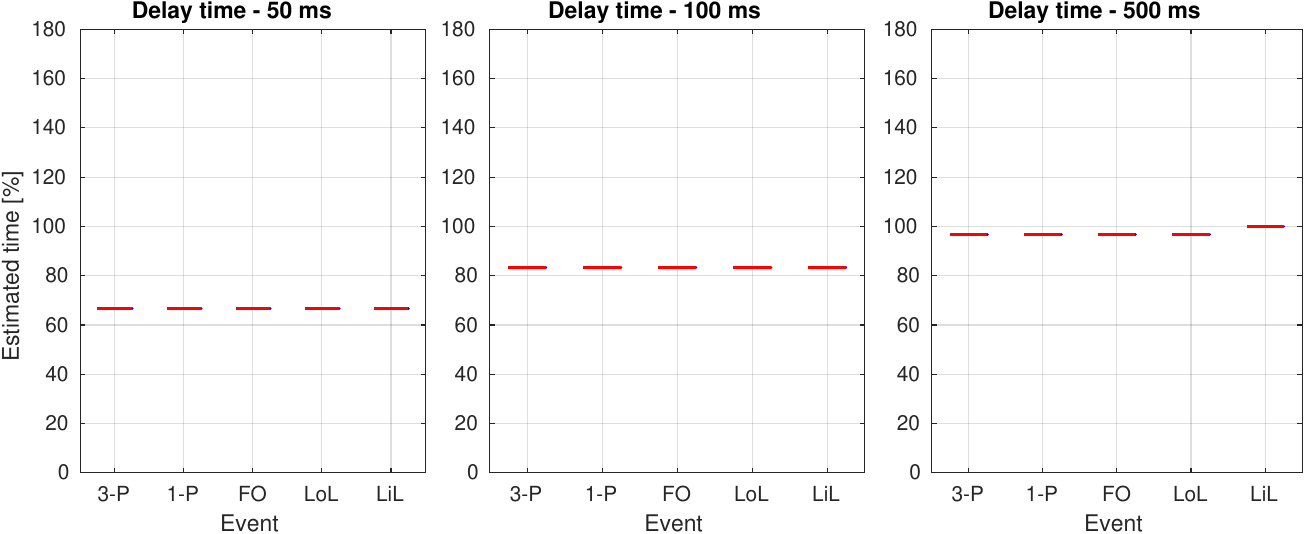}
         \label{fig:boxplot_TVE_01}
     \end{subfigure}
     \begin{subfigure}[b]{1\textwidth}
         \centering
         \vspace*{0.3cm}
         \caption{$TVE=1 \%$.}
				 \includegraphics[width=.9\columnwidth]{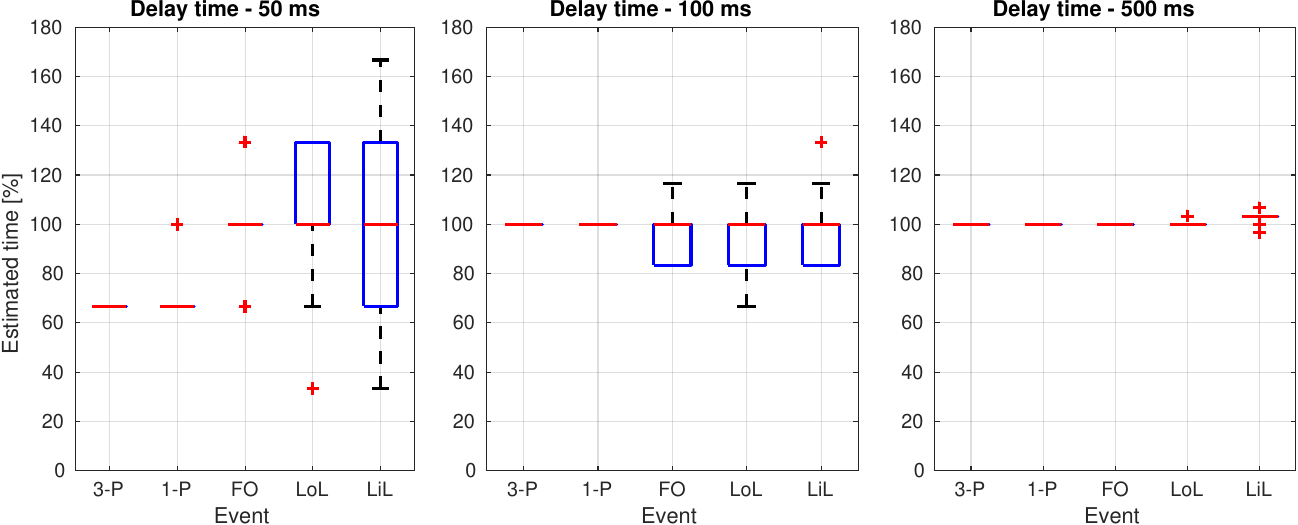}				
         \label{fig:boxplot_TVE_1}
     \end{subfigure}
     \begin{subfigure}[b]{1\textwidth}
         \centering
         \vspace*{0.3cm}
         \caption{$TVE=3 \%$.}
				 \includegraphics[width=.9\columnwidth]{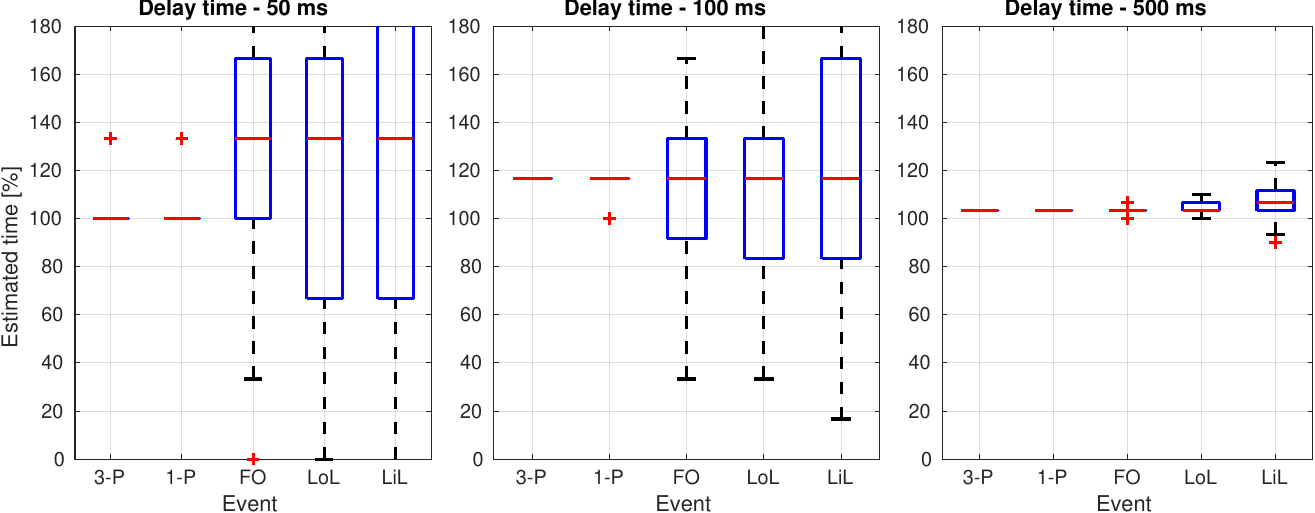}	 
         \label{fig:boxplot_TVE_3}
     \end{subfigure}
     \caption{Boxplot of estimated times in all cases.}
		 \label{fig:boxplot_TVE}
\end{figure}

Finally, all these cases were tested after 100 MonteCarlo trials. The results obtained are shown in Figs. \ref{fig:boxplot_TVE_01}-\ref{fig:boxplot_TVE_3}. The results show a good performance of the methodology especially when the TVE is 1 \% or less. When the TVE is 3 \%, highest value permitted by the synchrophasor standard \cite{IECStandard2018}, the results depends strongly of the dynamic event to be evaluated. It is clear that the 3-p and 1-P events are more suitable to enhance the performance of our proposal. On the other hand, in all cases it seems that the results improve as the communication delay increases. There is a big difference respect to \cite{Cui2023} that shows the opposite tendency. It should be notice that when the noise is low, even though the error seems to be high, the results differ only by one sample in time with no deviation form the mean value. It is important to mention that the value of the delay could be unknown and the methodology could still detect the difference in time between the signals because there is no \textit{a-priori} model about the value of the delay.

\section{Conclusions}
This paper presents a new methodology to detect scenarios where the communication link produces a time-shift of the reactive control signal for PMSG-based turbines. This new methodology, based on a dynamic state estimator, adopts a reduced model of the WT to obtain a real-time solution to estimate the delay communication. The estimation obtained can be used to correct the model simulated by transmission system operators to analyze dynamic simulations of the power system response.

Numerical results have shown that reactive control signal can be estimated accurately. We can conclude that accuracy in the estimation of the time delay depends strongly on the event to be evaluated. The reason is that the intensity of the reactive control signal determines the signal-to-noise ratio. Strong perturbations produce better results. However if the delay is high enough it could be analized using other dynamic events such as loss of loads, loss of lines or forced oscilations.

Finally, as an advantage of this methodology, we can state that it does not depend on the nature of the oscillation produced by the delay. On the other hand, this methodology can be extended and applied to other renewable sources, including solar plants. However, complex power generation systems with control signals far from the PCC are more difficult to estimate. Further studies should be conducted to determine if system observability can be reached.

%\section*{Appendix}
%%\section{Nomenclature} \label{sec:nomenclature}
%%The nomenclature used for the Kalman filter algorithm is listed in Table \ref{tab:nomenclature1}. All the variables and constants used throughout this article are listed in Table \ref{tab:nomenclature2}.
%%They are assessed using the per-unit system, unless they are time constants, which
%%are quantized in seconds, or phases, in radians. 
%\section{Algebraic equations} \label{sec:algebraic_equations}
%To complete the set of equations and to implement the UKF a set of algebraic equations from the LCL filter is necessary:
%
%\begin{subequations}
%\begin{align}
%q_{filter} &= v_{gd} \, i_{gq} - v_{gq} \, i_{gd}
%\end{align}
%\end{subequations}

%\section{Acknowledgments}
%The work of P. Marchi was supported by a CONICET Posdoctoral fellowship. This work was partially funded by the UREE 4 FONARSEC project: ``Development of Synchrophasor Measurements for Smart Electrical Grids''. The FONARSEC is funded by the Ministry of Science, Technology and Innovaton of Argentina.

\bibliographystyle{elsarticle-num-names}
\bibliography{biblio}

\end{document}